\documentclass[NumberedRefs,reprint,turnofflinenumbers]{JASA-El}

\usepackage{siunitx}
\usepackage[capitalise]{cleveref}
\usepackage{amssymb,amsmath}
\usepackage{mathtools}

\newcommand{\funcname}[1]{\mathrm{#1}}
\newcommand{\epow}[1]{\funcname{e}^{#1}}
\renewcommand{\jmath}{\mathrm{j}}

\newcommand{\sampidx}{n} 
\newcommand{\frameidx}{\ell} 
\newcommand{\freqidx}{k} 
\newcommand{\sampfreq}{f_s}

\newcommand{\framelen}{M}
\newcommand{\frameshift}{H}
\newcommand{\overlapratio}{R}
\newcommand{\numframes}{L}
\newcommand{\numbins}{K}

\newcommand{\framelenT}{\framelen_t}

\newcommand{\cleansig}{s}

\newcommand{\noisysig}{x}
\newcommand{\windowsig}{w}

\newcommand{\cleansigF}{S}
\newcommand{\noisesigF}{V}
\newcommand{\noisysigF}{X}

\newcommand{\sigT}[2]{\ensuremath{#1(#2)}} 
\newcommand{\noisyT}[1]{\sigT{\noisysig}{#1}} 
\newcommand{\windowT}[1]{\sigT{\windowsig}{#1}} 

\newcommand{\noisyTc}{\noisyT{\sampidx}} 
\newcommand{\windowTc}{\windowT{\sampidx}} 

\newcommand{\sigTF}[3]{\ensuremath{#1_{#2,#3}}} 

\newcommand{\noisyTF}[2]{\sigTF{\noisysigF}{#1}{#2}}

\newcommand{\noisyTFc}{\noisyTF{\freqidx}{\frameidx}} 

\newcommand{\C}{\mathbb{C}}

\newcommand{\Cmat}[2]{\C^{#1\minitimes#2}}

\DeclarePairedDelimiter{\abs}{\lvert}{\rvert}
\newcommand{\phase}[1]{\phi_{#1}}
\newcommand{\estmag}[1]{\abs{\widehat{#1}}}
\newcommand{\estphase}[1]{\widehat{\phi}_{#1}}

\newcommand{\eqcomma}{\,,} 
\newcommand{\eqperiod}{\,.} 
\newcommand{\minitimes}{{\mkern-2mu\times\mkern-2mu}}
\newcommand{\citenr}[1]{Ref.~\citen{#1}}

\begin{document}

\title[Phase-Aware Deep Speech Enhancement: It's All About The Frame Length]{Phase-Aware Deep Speech Enhancement: It's All About The Frame Length}

\author{Tal Peer}
\email{tal.peer@uni-hamburg.de}\correspondingauthor
\author{Timo Gerkmann}
\email{timo.gerkmann@uni-hamburg.de}
\affiliation{Signal Processing (SP), Universität Hamburg, Germany}

\date{\today} 

\begin{abstract}
    Algorithmic latency in speech processing is dominated by the frame length
    used for Fourier analysis, which in turn limits the achievable performance
    of magnitude-centric approaches. As previous studies suggest the importance
    of phase grows with decreasing frame length, this work presents a
    systematical study on the contribution of phase and magnitude in modern Deep
    Neural Network (DNN)-based speech enhancement at different frame lengths.
    Results indicate that DNNs can successfully estimate phase when using short
    frames, with similar or better overall performance compared to using longer
    frames. Thus, interestingly, modern phase-aware DNNs allow for low-latency
    speech enhancement at high quality.
\end{abstract}

\maketitle

\section{Introduction}
\label{sec:intro}
Single-channel speech enhancement is often carried out in the time-frequency
domain where the signals are represented by their time-varying frequency
content. To obtain the time-frequency representation, one applies a
transformation such as the short-time Fourier transform (STFT) which has a
number of free parameters. These parameters (namely frame length, frame shift
and window function, see also \cref{sec:pre}) must be chosen appropriately, e.g.
based on physical characteristics of the speech signal. However, not only the
signal should be considered, but also the algorithm which will be applied on the
time-frequency representation; the choice of STFT parameters should result in a
representation that is the most \emph{useful} for the algorithm at hand.\cite{virtanenTimeFrequencyProcessingSpectral2018}

In this letter, we focus expressly on the choice of frame length for speech
enhancement algorithms based on deep neural networks (DNNs) and specifically
consider phase-aware approaches. Choosing an adequate frame length is an
important decision in the design of STFT-based systems. On one hand, it largely
determines the overall algorithmic latency of the system, hence real-time
systems such as hearing aids would tend to use short frames. On the other hand,
short frames lead to limited spectral resolution which hinders many algorithms and also result in a larger number of frames for a given signal, potentially increasing computational complexity, e.g. when using temporal convolutions.
The STFT representation is complex-valued and commonly separated into a
\emph{magnitude spectrogram} and \emph{phase spectrogram}. The relevance of the
phase spectrogram to the speech enhancement task has been a topic of debate.
Traditionally it has been considered to be of little to no importance due to
empirical studies \cite{wangUnimportancePhaseSpeech1982} as well as theoretical
results.\cite{ephraimSpeechEnhancementUsing1984} However, more recent studies
have shown that phase does carry speech-relevant
information.\cite{paliwalImportancePhaseSpeech2011a,gerkmannPhaseProcessingSingleChannel2015}
Motivated by these findings, phase-aware speech processing has been enjoying a
certain renaissance and several phase-aware methods have been
proposed.\cite{lerouxConsistentWienerFiltering2013,gerkmannMMSEoptimalEnhancementComplex2014a,gerkmannBayesianEstimationClean2014b,krawczykSTFTPhaseReconstruction2014b,mowlaeeHarmonicPhaseEstimation2015}

In recent years DNNs have rapidly become the tool of choice in many fields,
including audio and speech processing. Consequently, many recent phase-aware
speech enhancement and source separation methods use a DNN to either directly
estimate the phase spectrogram
\cite{takahashiPhaseNetDiscretizedPhase2018,afourasConversationDeepAudioVisual2018,lerouxPhasebookFriendsLeveraging2019}
or estimate phase derivatives and reconstruct the phase from
them.\cite{zhengPhaseAwareSpeechEnhancement2019a,masuyamaPhaseReconstructionBased2020}
Other DNN-based approaches include directly operating on complex spectrograms
without separating into magnitude and phase
\cite{williamsonComplexRatioMasking2016,tanLearningComplexSpectral2020,huDCCRNDeepComplex2020a}
or simply taking phase into consideration for improved magnitude
estimation.\cite{erdoganPhasesensitiveRecognitionboostedSpeech2015}

Some authors have taken a different path and altogether replaced the STFT-based
representation with a learned encoder-decoder mechanism, which usually results
in a real-valued
representation.\cite{luoConvTasNetSurpassingIdeal2019,luoDualPathRnnEfficient2020,subakanAttentionAllYou2021}
An interesting aspect of these learned encoder-decoder approaches is that they
show very good performance when using very short frames of about \SI{2}{\ms},
even going as short as \SI{0.125}{\ms}.\cite{luoDualPathRnnEfficient2020} This
stands in sharp contrast to traditional STFT-based approaches which generally
use frame lengths of about \SIrange{20}{60}{\ms}. Note that while learned
encoder-decoder approaches have been originally proposed for source separation,
they also show good performance on the speech enhancement task.
\cite{koyamaExploringBestLoss2020,wangCompensationMagnitudePhase2021}

Following the publication of the pioneering learned encoder-decoder Conv-TasNet
model,\cite{luoConvTasNetSurpassingIdeal2019} several authors have proposed
extensions and analyses. Among other results, it has been shown that the main
contributing factors to the performance of Conv-TasNet are the use of short
frames and time-domain loss function, not the learned
encoder-decoder.\cite{ditterMultiPhaseGammatoneFilterbank2020,heitkaemperDemystifyingTasNetDissecting2020}
It has also been shown that when replacing the learned encoder with the STFT,
the optimal set of input features depends on the chosen frame
length;\cite{parienteFilterbankDesignEndtoend2020,heitkaemperDemystifyingTasNetDissecting2020}
for longer frames (\SIrange{25}{64}{\ms}) the magnitude spectrum works well,
while shorter frames (\SIrange{2}{4}{\ms}) show better performance only with
the full complex spectrogram as input (in form of concatenated real and
imaginary parts). This observation is especially important, since it means that
phase-aware speech processing (with either implicit or explicit phase
estimation) should possibly employ different frame lengths than magnitude-only
processing. 

\begin{figure}
    \centering
    \includegraphics[width=0.6\textwidth]{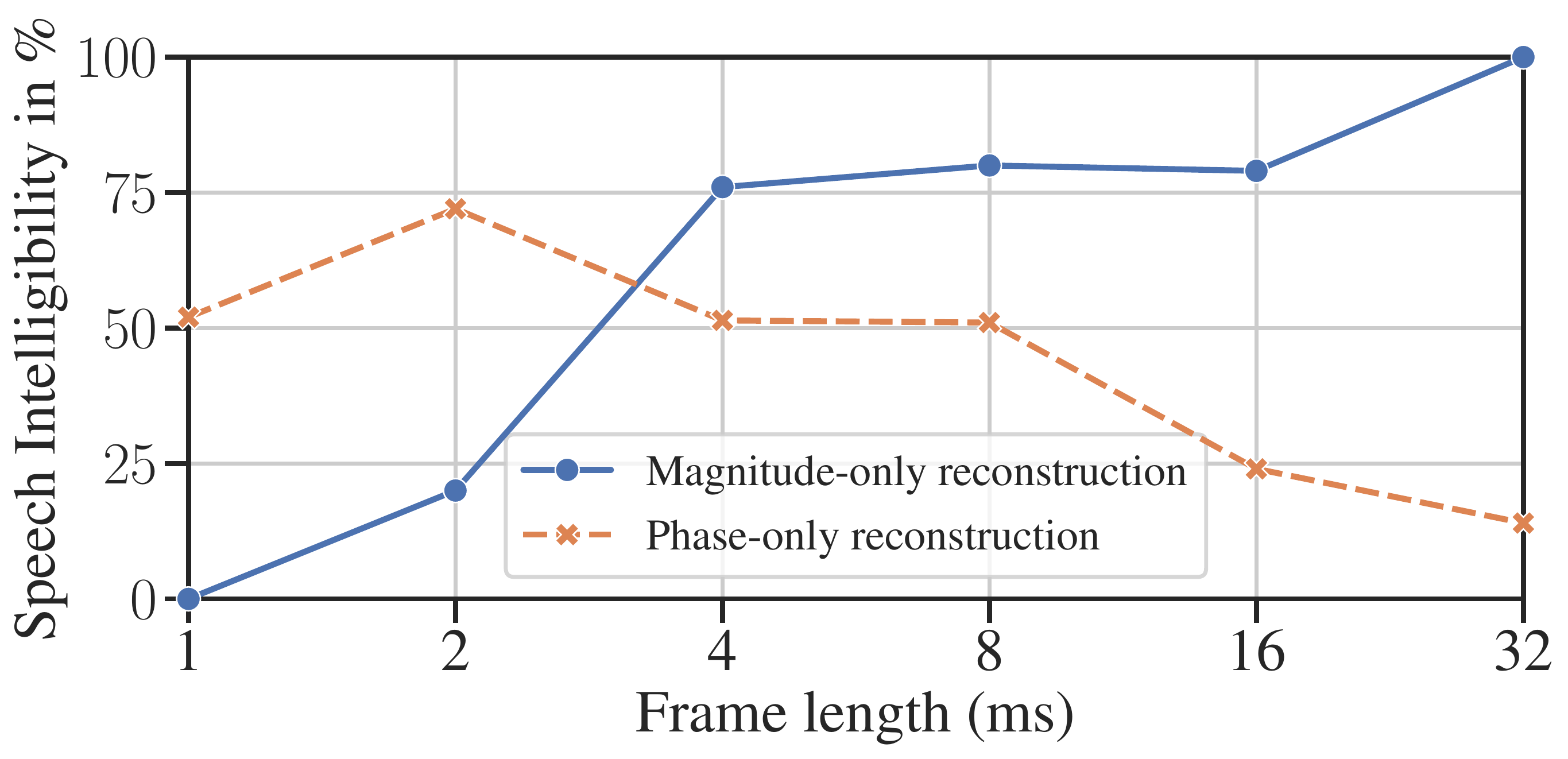}
    \caption{Redrawn excerpt of experimental results by Kazama et
    al.\cite{kazama2010significance} (shown here for illustration purposes).
    When reconstructing speech either from the magnitude spectrogram or phase
    spectrogram (replacing the other component by noise), the intelligibility of
    the resulting signal strongly depends on the choice of frame length.}
    \label{fig:kazama}
\end{figure}

While the choice of loss function in phase-aware speech enhancement DNNs has
been studied with respect to perceptual
measures,\cite{wangCompensationMagnitudePhase2021} and the effect of STFT parameters on magnitude-only DNNs has also been analyzed,\cite{takeuchiEffectSpectrogramResolution2020} we are not aware of an
analysis regarding the choice of frame length in the phase-aware setting. Previous studies unrelated to
DNNs have shown that the importance of phase to speech-related tasks varies with
the choice of STFT parameters. In particular, it has been shown that departing
from the typical frame lengths used in speech processing (corresponding to about
\SIrange{20}{40}{\ms}) and either using shorter frames
\cite{kazama2010significance,peerIntelligibilityPredictionSpeech2021} or using a
window shape that effectively shortens the frame
\cite{paliwalImportancePhaseSpeech2011a} can result in very good signal
reconstruction from the phase spectrogram alone. A similar result has been
observed for longer-than-typical
frames.\cite{alsterisImportanceWindowShape2004,kazama2010significance} However,
as long frames yield algorithmic latencies which may be prohibitive for many
real-time speech processing devices, here we choose to focus on short frames for
their potential benefit regarding latency. \Cref{fig:kazama}, based on results from
\citenr{kazama2010significance}, shows how the contribution of phase and
magnitude to the intelligibility of reconstructed clean speech changes with
varying frame length. One observes that as the frames become shorter, phase
becomes more important while magnitude gradually loses relevance. Note, however,
that these findings are based on somewhat artificial signal reconstruction
experiments on oracle data --- whether or not they also apply to actual speech
enhancement, source separation, etc. remains unclear.

Typical frame lengths in speech processing --- around \SI{32}{\ms} ---
correspond to an interval that is short enough to be considered quasi-stationary
but long enough to cover multiple fundamental periods of voiced speech (whose
fundamental period lies between \SI{2}{\ms} and
\SI{12.5}{\ms}).\cite{paliwal2010preference} These considerations apply to the
magnitude spectrogram but not necessarily to the phase spectrogram. Indeed, it
seems that the irrelevance often attributed to the phase spectrogram is in part
due to the choice of frame length in experiments. Note that existing model-based
phase estimation methods do not typically operate on short
frames\cite{krawczykSTFTPhaseReconstruction2014b,mowlaeeHarmonicPhaseEstimation2015,magronPhaseReconstructionSpectrograms2015,peerPlosiveEnhancementUsing2021b}
and thus do not attempt to take advantage of the findings regarding phase
importance and frame length in perceptual studies.

Based on these previous results and observations, we seek to answer two
questions in this letter: a) How does the choice of frame length affect
magnitude and phase estimation in a phase-aware DNN? b) Does DNN-based phase
estimation allow the use of shorter frames, thus reducing algorithmic latency?
Using an example DNN with explicit phase estimation, we analyze and compare the
performance under different frame lengths. In order to gain further insight, we
also attempt to characterize the relative contribution of the magnitude and
phase spectrograms at each frame length and show that the aforementioned
observations on the importance of phase in short frames also carry over to the
context of speech enhancement.
\section{Preliminaries}
\label{sec:pre}
The STFT of a discrete time-domain signal $\noisyTc$ is computed by segmenting
the signal into overlapping frames of length $\framelen$ and shift
$\frameshift$. A real-valued multiplicative window function $\windowTc$ is
applied to each frame, which is then transformed to the frequency domain with
the discrete Fourier transform (DFT). Assuming the one-sided $\framelen$-point
DFT is used, we obtain the \emph{complex spectrogram} $\noisysigF \in
\Cmat{\numbins}{\numframes}$, defined as
\begin{equation}
    \noisyTFc = \sum_{\sampidx=0}^{\framelen-1} \noisyT{\frameidx \frameshift + \sampidx} \windowTc \epow{-\jmath 2\pi \frac{\freqidx \sampidx}{\framelen}} \eqcomma 
\end{equation}
where $\freqidx$ is the frequency index, $\frameidx$ is the frame index,
$\numbins = \frac{M}{2}+1$ is the number of frequency bins and $\numframes$ is
the number of time frames. Unless otherwise noted, we always consider the whole
spectrogram and thus omit the indices in the following. We also define an
\emph{overlap ratio} $\overlapratio = \frac{\framelen - \frameshift}{\framelen}$
for convenience. As $\framelen$ is the number of samples in a single frame, we
define $\framelenT = \frac{\framelen}{\sampfreq}$ (where $\sampfreq$ is the
sampling frequency) as the physical frame length, measured in seconds. The term
\emph{frame length} will refer to $\framelenT$ from this point onwards.

In the context of speech enhancement we consider an additive noise model, which
is expressed in polar coordinates in terms of the magnitude spectrogram
$\abs{\noisysigF}$ and phase spectrogram $\phase{\noisysigF}$:
\begin{equation}
    \abs{\noisysigF} \epow{\jmath \phase{\noisysigF}} = \abs{\cleansigF} \epow{\jmath \phase{\cleansigF}} + \abs{\noisesigF} \epow{\jmath \phase{\noisesigF}} \eqcomma
\end{equation}
where $\cleansigF$ and $\noisesigF$ are the clean speech signal and an additive
noise component, respectively. Given the noisy signal $\noisysigF$, the task is
to compute an estimate $\widehat{\cleansigF} = \estmag{\cleansigF} \epow{\jmath
\estphase{\cleansigF}}$ which is subsequently transformed back into the
time-domain, yielding the estimated clean signal
$\widehat{\cleansig}(\sampidx)$.
\section{Neural network architecture}
\label{sec:dnn}
\begin{figure}[ht]
    \includegraphics[width=\textwidth]{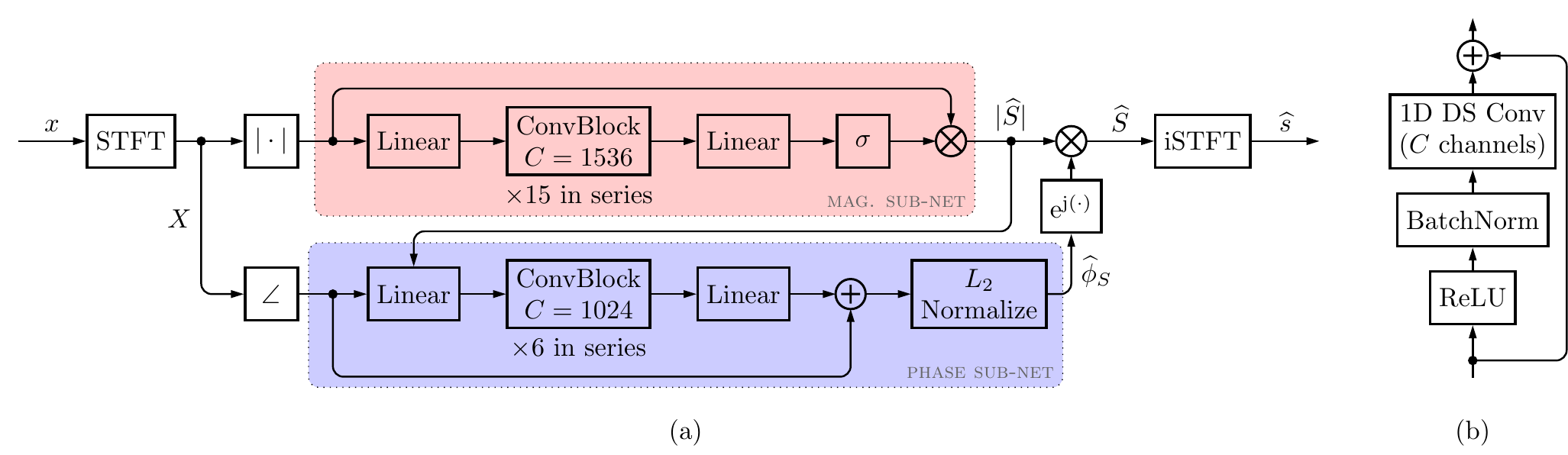}
    \caption{(a) Overview of the proposed network architecture, based on the
    audio-visual model by Afouras et
    al.,\cite{afourasConversationDeepAudioVisual2018} albeit only using the
    noisy speech signal as input. Note that the input to the phase sub-network
    consists of the estimated magnitude along with the cosine and sine of the
    noisy phase, which are shown here as a single input for simplicity. (b)
    Detail of the basic convolutional block.}
    \label{fig:network}
\end{figure}

The DNN architecture proposed here is an adaptation of a previously proposed
model\cite{afourasConversationDeepAudioVisual2018} for audio-visual speech
separation and enhancement, consisting of loosely coupled magnitude and phase
sub-networks. Although we do not consider an audio-visual input here, this model
is relatively simple and performs explicit estimation of both magnitude and
phase, which is essential for our experiments. The parts pertaining to the video
stream are simply omitted and the model is adapted accordingly. The resulting
network is depicted in \cref{fig:network} and described below.

Both sub-networks are realized as convolutional neural networks (CNNs) using
one-dimensional depthwise separable convolution layers
\cite{cholletXceptionDeepLearning2017} along the time axis (the different
frequency bins at the input are considered as channels in this setup). Both
networks consist of multiple identical residual blocks; The basic building block
is composed of a pre-activation (ReLU), a batch normalization layer and a
convolutional layer whose output is added to the block's input.
\subsection{Magnitude sub-network}
The magnitude sub-network takes the noisy magnitude spectrogram
$\abs{\noisysigF}$ outputs a real mask which is applied to the noisy STFT
magnitude spectrogram to produce a magnitude estimate (note that the original
network\cite{afourasConversationDeepAudioVisual2018} used the mel-scale
spectrogram along with video features as input). The noisy magnitude spectrogram
is fed through a chain of 15 convolutional blocks with 1536 input/output
channels each. Linear layers at the input and output help to model the
inter-frequency relationships and project the data into the correct dimensions.
A sigmoid activation function is applied to the output, resulting in a real mask
with values in $[0,1]$.  The real mask is multiplied with the input, resulting
in a magnitude estimate $\estmag{\cleansigF}$.
\subsection{Phase sub-network}
The input to the phase sub-network is a concatenation of $\estmag{S}$,
$\cos(\phase{\noisysigF})$ and $\sin(\phase{\noisysigF})$ along the frequency
axis. This is fed into a linear input layer and subsequently through 6
convolutional blocks with 1024 channels and a linear output layer. The output of
the linear layer is treated as the concatenated cosine and sine of the phase
residual, which are added to the respective inputs. The resulting estimate is
$L_2$-normalized to ensure that the cosine and sine outputs are consistent with
each other (i.e. that they represent a unit vector on the complex plane). By having the estimated magnitude as an additional input, the phase sub-network can learn to focus on high-energy spectrogram regions, resulting in an overall better phase estimate.
\subsection{Training procedure}
The magnitude and phase sub-networks are trained jointly using a training set
consisting of pairs of noisy and clean speech samples at different
signal-to-noise ratios (SNR), see \cref{sec:exp_data} for further details. We
employ a time-domain loss, namely the negative scale invariant signal to
distortion ratio (SI-SDR).\cite{lerouxSDRHalfbakedWell2019} The SI-SDR loss has
been shown to produce superior results over frequency-domain loss functions when
both magnitude and phase spectrograms are
estimated.\cite{heitkaemperDemystifyingTasNetDissecting2020}
\section{Experiments}
\label{sec:exp}
The main experiment we conduct is a comparison of the model's performance under
variation of the STFT frame length $\framelenT$, in terms of perceptual
measures. Since the model we consider includes explicit estimation of phase and
magnitude, we are able to also analyze and quantify the relative contribution of
magnitude and phase estimation, again as a function of frame length. This
analysis is conducted in a manner comparable with the perceptual experiments in
previous
works,\cite{paliwalImportancePhaseSpeech2011a,kazama2010significance,peerIntelligibilityPredictionSpeech2021}
although here we use estimates of the clean magnitude and phase, rather than the
clean or noisy signals. For each frame length, we produce three estimates of the
clean speech signal --- the actual output of the network as well as two
synthetic signals composed of the estimated magnitude and noisy phase or
vice-versa:
\begin{equation}\label{eq:est_sigs}
        \widehat{\cleansig} = \mathrm{iSTFT}\{\estmag{\cleansigF} \epow{\jmath \estphase{\cleansigF}}\} \eqcomma \quad 
        \widehat{\cleansig}_{\mathrm{mag}} = \mathrm{iSTFT}\{\estmag{\cleansigF} \epow{\jmath \phase{\noisysigF}}\} \eqcomma \quad
        \widehat{\cleansig}_{\mathrm{ph}} = \mathrm{iSTFT}\{\abs{\noisysigF} \epow{\jmath \estphase{\cleansigF}}\} \eqperiod 
\end{equation}
To allow for a fair comparison we must keep the number of DNN parameters
constant. In the case of the network architecture we consider, the number of
parameters depends on the number of frequency bins $\numbins$. Hence, we
zero-pad the frames prior to applying the DFT, resulting in a constant number of
bins $\numbins = 257$, which corresponds to the longest frames we consider
($\framelenT = \SI{32}{\ms}$) at $\sampfreq = \SI{16}{\kHz}$. All experiments
employ a square-root Hann window with an overlap ratio $\overlapratio =
\frac{1}{2}$. The same window is used for the forward and inverse STFT
operations.
\subsection{Data and training details}
\label{sec:exp_data}
For training we use clean and noisy excerpts from the 2020 Deep Noise
Suppression (DNS) dataset \cite{reddyINTERSPEECH2020Deep2020} with SNR $\in
\{-5, 0, \dotsc, 10\}\si{\decibel}$. Each excerpt is \SI{2}{\s} long and the
data set contains in total \SI{100}{\hour} of speech, from which
\SI{80}{\percent} are used for training and the remaining \SI{20}{\percent}
for validation. We train all models using the Adam optimizer, a batch size of 32
and a learning rate of \num[scientific-notation =
true,retain-unity-mantissa=false]{e-4}. Training is stopped if the validation
loss has not decreased for 10 epochs.

Evaluation is performed on a custom test set composed of clean speech from the
WSJ corpus \cite{paulDesignWallStreet1992} and noise from the CHiME3
dataset,\cite{barker2015third} mixed with SNR $\in \{-10,-5,\dotsc,
20\}\si{\decibel}$. This test set contains 672 excerpts in total. All training
and evaluation data is sampled at $\sampfreq = \SI{16}{\kilo\hertz}$. 

\subsection{Evaluation details}
In addition to evaluation in terms of instrumental quality and intelligibility
measures (POLQA, ESTOI) using the entire evaluation dataset, we also report the
results of a small-scale listening experiment in which participants were asked
to rate the overall quality of the estimated signals introduced in
\cref{eq:est_sigs}. The listening experiment consisted of 12 trials. In each
trial the participants were presented with a reference clean signal from the WSJ
corpus and then asked to rate the quality of eight different signals: A noisy
version from the evaluation dataset (at \SI{0}{\decibel} SNR), the reference
itself, the model output $\widehat{\cleansig}$, and the phase/magnitude-based
reconstructions ($\widehat{\cleansig}_{\mathrm{ph}}$ and
$\widehat{\cleansig}_{\mathrm{mag}}$) for two representative frame lengths ---
\SI{32}{\ms} and \SI{4}{\ms}. Ten normal-hearing individuals aged 25 to 43
participated in the experiment and rated each signal on a continuous quality
scale (CQS) from 0 to 100.
\section{Results and discussion}
\label{sec:results}
\begin{figure}[ht]
    \includegraphics[width=\textwidth]{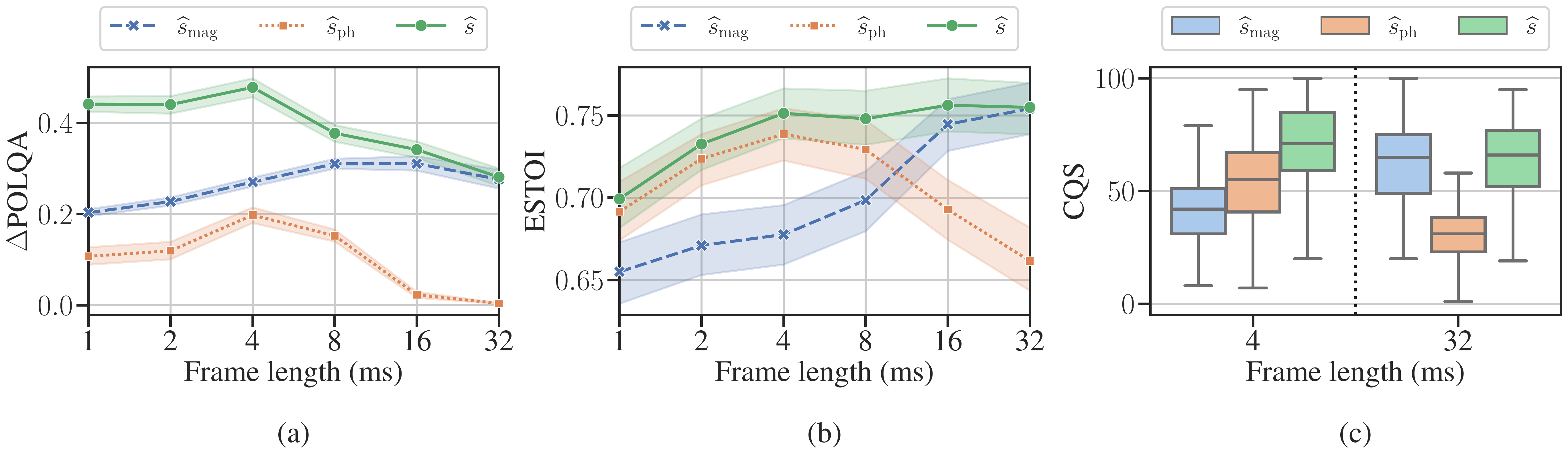}
    \caption{(a),(b) Evaluation results on the WSJ/CHiME test set (see
    \cref{sec:exp_data}). The plots show the mean improvement of POLQA score and
    the mean ESTOI score over all input SNRs as function of the frame length.
    Error bands depict the \SI{95}{\percent} confidence interval. (c) Results
    of listening experiment on two frame lengths, where participants were asked
    to rate the different signals on a continuous quality scale (CQS) in the
    range 0--100.}\label{fig:chime_results}
\end{figure}
Evaluation results are depicted in \cref{fig:chime_results}. We first
consider the effect of frame length on the overall estimate
$\widehat{\cleansig}$, which uses both the estimated magnitude and estimated
phase for iSTFT reconstruction. Intelligibility (in terms of ESTOI) is not
affected much by the choice of frame length except for very short frames, which
cause rapid degradation. In terms of speech quality ($\Delta$POLQA), however, we
observe a consistent improvement with decreasing frame length. The improvement
in speech quality reaches a maximum at $\framelenT = \SI{4}{\ms}$, after which
it starts to decline, while still reaching relatively high values for very short
frames of \SIrange{1}{2}{\ms}. 

For both POLQA and ESTOI, the magnitude-based and phase-based estimates
($\widehat{\cleansig}_{\mathrm{mag}}$ and $\widehat{\cleansig}_{\mathrm{ph}}$,
respectively) show an interesting picture: At $\framelenT = \SI{32}{\ms}$,
$\widehat{\cleansig}_{\mathrm{mag}}$ reaches similar values to
$\widehat{\cleansig}$ and the phase-based estimate
$\widehat{\cleansig}_{\mathrm{ph}}$ shows virtually no improvement over the
noisy input. As the frame length decreases this gradually changes: The
magnitude-based estimate loses quality and intelligibility, while the opposite
holds for the phase-based estimate. Note that although the magnitude-based
estimate shows degraded quality for short frames, the phase spectrum's
contribution is still sufficient to maintain and even boost overall performance
of the joint estimate. We attribute this behavior to the relative contribution
of phase and magnitude spectra and the interplay between them.

Results of the listening experiment largely agree with the instrumental
measures. The joint estimate $\widehat{\cleansig}$ received very similar quality
scores for short and long frames (however without the rising trend shown by
POLQA). This indicates, again, that the use of shorter frames (e.g. for reduced
latency) does not imply a sacrifice in terms of enhancement performance. The
scores given to the phase and magnitude-based estimates are consistent with
POLQA and ESTOI scores and support the proposition, that the importance of phase
and magnitude estimation is dependent on the chosen frame length. In particular,
at the short frame regime, phase estimation plays a larger role than magnitude
estimation.

\begin{figure}[ht]
    \includegraphics[width=\textwidth]{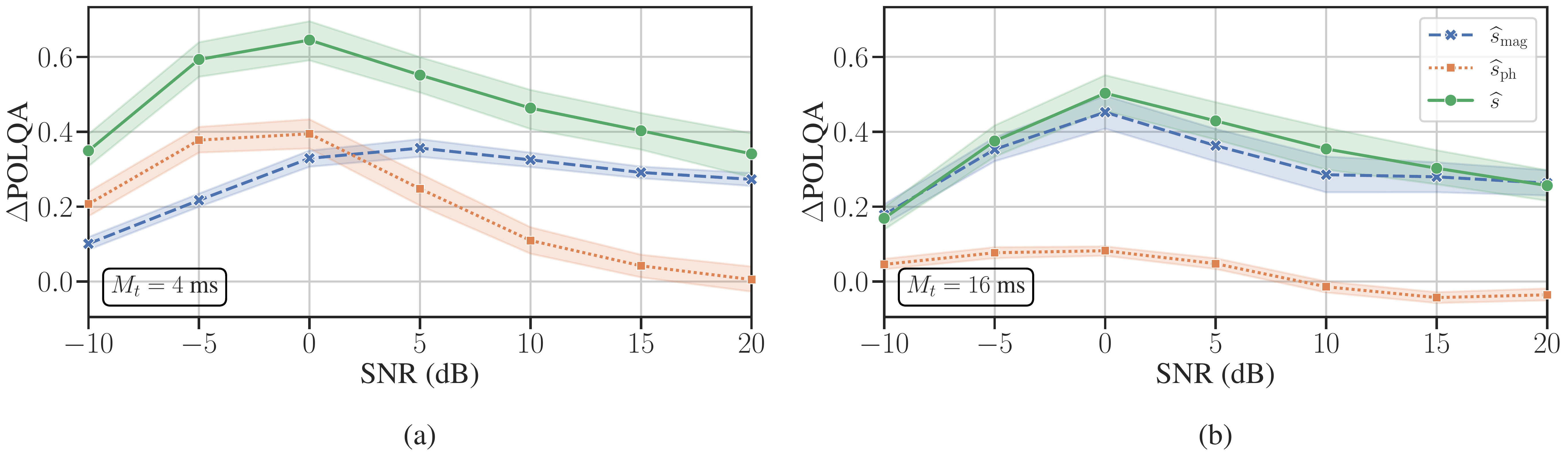}
    \caption{Mean POLQA improvement on the WSJ/CHiME test set for $\framelenT
    \in \{4,16\}\si{\ms}$, shown as a function of input SNR.}
    \label{fig:chime_snr}
\end{figure}

Since the results in \cref{fig:chime_results} are averaged across all SNRs, we
provide further insight in \cref{fig:chime_snr} by showing the POLQA improvement
at different SNRs for two selected frame lengths (\SI{4}{\ms}, \SI{16}{\ms}).
Besides the overall better performance of the shorter frame length and the
unimportance of phase estimation with long frames (cf.
\cref{fig:chime_results}), the quality of magnitude-based and phase-based
estimates shows much stronger dependence on SNR. In particular, the phase-based
estimate actually surpasses the magnitude-based estimate in terms of POLQA at
low SNRs ($\leq \SI{0}{\decibel}$), suggesting that phase estimation is
especially beneficial at difficult noise conditions, in accordance with previous
perceptual studies.\cite{krawczyk-beckerEvaluationPerceptualQuality2016a} This
also translates into the joint estimation case (i.e. $\widehat{\cleansig}$),
where the difference in $\Delta$POLQA between the frame lengths is more
pronounced at low SNRs.
\section{Conclusion}
In this work we have presented a study on the effect of frame length in
STFT-based phase-aware speech enhancement with DNNs. Results indicate that the
use of short frames in this context (down to \SI{4}{\ms}) does yield similar or
better performance in terms of instrumental and subjective quality measures
compared to long frames (\SI{32}{\ms}). Furthermore, by explicitly estimating
phase and magnitude we are able to show that varying the frame length affects
the individual contribution of magnitude and phase estimation to the quality of
the combined output --- at short frames, it is dominated by phase estimation.
These findings suggest that by employing explicit phase estimation, speech
enhancement DNNs can achieve reduced latency, which is of great importance for
speech communication devices such as hearing aids or virtual assistants.
\begin{acknowledgments}
This work has been funded by the Deutsche Forschungsgemeinschaft (DFG, German
Research Foundation) -- project number \texttt{247465126}. We would like to
thank J. Berger and Rohde\&Schwarz SwissQual AG for their support with POLQA.
\end{acknowledgments}
\bibliography{refs}
\end{document}